\documentclass[fleqn,usenatbib]{mnras}

\usepackage{newtxtext,newtxmath}

\usepackage[T1]{fontenc}
\usepackage{aecompl} 
\DeclareRobustCommand{\VAN}[3]{#2}
\let\VANthebibliography\thebibliography
\def\thebibliography{\DeclareRobustCommand{\VAN}[3]{##3}\VANthebibliography}

\usepackage{graphicx}	
\usepackage{amsmath}	

\usepackage{booktabs} %
\usepackage{multirow} 
\usepackage{threeparttable} %
\usepackage{color, xcolor}
  \usepackage{graphicx} 
  \usepackage{epstopdf}
\usepackage{subfigure}
\usepackage{float}
\usepackage{verbatim}
\usepackage{stfloats}
\usepackage{hyperref}
\usepackage{pgfplots}
\usepackage{lineno}
\DeclareUnicodeCharacter{2212}{\ensuremath{-}}

\title[Reflection of MAXI J1348-630]{The spin measurement of MAXI J1348-630 using the Insight-HXMT data}

\author[Yujia Song et al.]{
Yujia Song,$^{1,2}$
Nan Jia,$^{1,2}$
Jun Yang$^{1,2}$
Ye Feng,$^{1,2}$
Lijun Gou$^{1,2}$\thanks{E-mail:
lgou@nao.cas.cn}
and Tianhua Lu$^{1,2}$
\\
$^{1}$Key Laboratory for Computational Astrophysics, National Astronomical Observatories, Chinese Academy of Sciences, Datun Road A20, Beijing 100012, China\\
$^{2}$School of Astronomy and Space Sciences, University of Chinese Academy of Sciences, Datun Road A20, Beijing 100049, China
}

\date{Accepted XXX. Received YYY; in original form ZZZ}

\pubyear{2023}

\begin{document}


\label{firstpage}
\pagerange{\pageref{firstpage}--\pageref{lastpage}}
\maketitle
\begin{abstract}
We report the results of fitting \textit{Insight-HXMT} data to the black hole X-ray binary MAXI J1348-430, which was discovered on January 26th, 2019, with the Gas Slit Camera (GSC) on-board \textit{MAXI}. Several observations at the beginning of the first burst were selected, with a total of 10 spectra. From the residuals of fits using disk plus power law models, X-ray reflection signatures were clearly visible in some of these observations. We use the state-of-the-art $\mathtt{relxill}$ series reflection model to fit  six spectra with distinct reflection signatures and a joint fit to these spectra. In particular, we focus on the results for the black hole spin values. Assuming $R_{\mathrm{in}} = R_{\mathrm{ISCO}}$, the spin parameter is constrained to be $0.82^{+0.04}_{-0.03}$ with 90\% confidence level (statistical only). 
\end{abstract}




\section{Introduction}
 
An important area of astrophysics is the X-ray binary, which consists of a compact object and a donor star. They can be divided into low-mass X-ray binaries (LMXBs, \citealt{sazonovGalacticLMXBPopulation2020}) and high-mass X-ray binaries (HMXBs, \citealt{fortinCatalogueHighmassXray2023a}) according to the different masses of the secondary stars. Within the Milky Way, HMXBs are distributed along the galactic plane, and LMXBs are spatially distributed near the galactic plane and center \citep{tanHighMassXrayBinary2021a}. Transient X-ray binaries remain dormant for long periods of time and occasionally have outbursts that typically last for weeks to months. The vast majority of black hole X-ray binaries (BHXRBs) in the Milky Way have been found as transient sources in LMXBs, with changes in X-ray luminosity accompanied by variations in timing during the outburst \citep{mcclintockBlackHoleSpin2014}. At the beginning of the burst, the source is faint and has an exponential cutoff near 100 keV, characterized by a hard power-law(PL) component \citep{turollaPowerlawTailsDynamical2002}. This period is called "low/hard State"(LHS). The X-ray spectra of LHS can generally comprise a dominant PL non-thermal component and sometimes a thermal component. In this state, the disk was either not observed above 2 keV or it appeared much cooler and truncated from the black hole (BH, \citealt{remillardXRayPropertiesBlackHole2006}). In addition, intense radio emission can be observed in this state, and sometimes quasi-periodic oscillations (QPOs) can be present \citep{stellaCorrelationsQuasiperiodicOscillation1999a}. Gradually the contribution of the thermal component begins to increase and the spectrum becomes soft. The source evolves from an LHS to a high/soft state (HSS) through an intermediate state (IMS, \citealt{homanEvolutionBlackHole2005}). In the HSS, the spectrum is dominated by thermal component, the radio emission is weak and the X-ray flux slowly declines, and QPOs are usually not observed \citep{shakura_sunyaev_1973,laurentSPECTRALINDEXFUNCTION2011}.

In a realistic astrophysical environment, black holes can be described by only two basic parameters: spin and mass. Spin is an important indicator of the formation and evolution of black holes, so it is of great importance to determine the spin of black holes by observation. Since isolated black holes are difficult to detect, accreting BHXRBs are the appropriate laboratories to study the physics of black holes. One method often used to measure the spin of black holes is the continuum-fitting (CF) method, using this method we have obtained the spin of dozens of black holes \citep{fengSpectralAnalysisNew2022,kushwahaAstroSatMAXIView2021}. However, it needs some system parameters such as the mass of the black hole $M$, the distance of the source $D$, and the inclination angle $i$ of the accretion disk around the black hole \citep{zhangBlackHoleSpin1997}. Besides CF method, by modeling the reflection component, \citet{fabianXrayFluorescenceInner1989} first proposed a method to determine the spin of a black hole, known as the Fe K$\alpha$ fitting method. The advantage of this approach is that the spin can be better constrained without information on the binary parameters. \citet{tanakaGravitationallyRedshiftedEmission1995} were the first to measure the spin using this method. Both methods assume that the inner radius of the accretion disk extends to the innermost stable circular orbit (ISCO), which is simply related to the spin parameter $a_{*}$ \citep{bardeenRotatingBlackHoles1972}. In fact, there is strong observational and theoretical evidence that the disk is quite sharply truncated at ISCO \citep{tanakaBlackHoleBinaries1995,steinerCONSTANTINNERDISKRADIUS2010}. The X-ray reflection spectra are characterized by several components, such as a forest of soft X-ray lines, the broad Fe emission line (in the 6.4-6.97 keV band, depending on the ionization state), and a Compton hump at 20-40 keV from scattered continuum off the photoelectric edge of the iron on the low side and Compton down-scattering at high energy \citep{chakrabartiSpectralPropertiesAccretion1995}. This approach has been used to measure more than a dozen systems and has been cross-checked with the CF method: GRO J1655-40 \citep{shafeeEstimatingSpinStellarMass2005,millerLARMASSBLACKHOLE2009a}; GX339-4 \citep{kolehmainenLimitsSpinDetermination2010,parkerNuSTARSWIFTOBSERVATIONS2016}; XTE J1550-564 \citep{steinerSpinBlackHole2011b}; GRS 1915+105 \citep{millerNuSTARSPECTROSCOPYGRS2013,sreehariAstroSatViewGRS2020}; 4U 1630-472 \citep{kingDISKWINDRAPIDLY2014,pahariAstroSatChandraView2018}; Cyg X-1 \citep{parkerNuSTARSUZAKUOBSERVATIONS2015,gouCONFIRMATIONCONTINUUMFITTINGMETHOD2014};IGR J179091-3624 \citep{wangReflectionComponentAverage2018,raoWHYIGRJ170912012}; LMC X-3 \citep{janaNuSTARSwiftObservations2021,steinerLOWSPINBLACKHOLE2014}; LMC X-1 \citep{janaNuSTARSwiftObservations2021,gouDETERMINATIONSPINBLACK2009}; MAXI J1659-152 \citep{routRetrogradeSpinBlack2020,fengEstimatingSpinBlack2022a}. 

The transient X-ray source MAXI J1348-630 was detected by the \textit{MAXI}/GSC instrument on 26 January 2019 \citep{yatabeMAXIGSCDiscovery2019}. It is an LMXB and contains a black hole \citep{zhangNICERObservationsReveal2020}. \citet{tominagaDiscoveryBlackHole2020} reported the first half year of monitoring of MAXI J1348-630. The observations show that the source is characterized by a low disk temperature ($\sim 0.75$ keV at the maximum) and a high peak flux ($\sim 0.56\ \mathrm{Crab}$ in the 2-10 keV band) in HSS, compared to other bright black hole transient sources (the maximum disk temperature almost $>1$ keV, the peak flux $\sim 0.8\ \mathrm{Crab}$ in the 2-10 keV). Assuming a face-on disk around a non-spinning black hole, the source distance and black hole mass are estimated to be $D\approx4 \mathrm{kpc}$ and $M\sim7(D/4\mathrm{kpc})M_{\odot}$, respectively. And they found that if the black hole is spinning and the disk is tilted, the black hole will be more massive. \citet{chauhanMeasuringDistanceBlack2021} estimated the probable distance of $D=2.2^{+0.5}_{-0.6}\mathrm{kpc}$ by studying the $\mathrm{H_{\uppercase\expandafter{\romannumeral1}}}$ absorption spectra. Using the scaling technique for the correlation between $\Gamma$ and the normalization proportional to $\dot{M}$, \citet{titarchukMAXIJ13486302023} estimated the disk inclination $i=(65\pm7)^{\circ}$ and the black hole mass of $M=14.8\pm0.9M_{\odot}$. Using the Fe K$\alpha$ fitting method, \citet{jiaDetailedAnalysisReflection2022} used nine Nuclear Spectroscopic Telescope Array (\textit{NuSTAR}, \citealt{harrisonNUCLEARSPECTROSCOPICESCOPE2013}) observations for a detailed spectral analysis of this source and they obtain the spin parameter $a_{*}=0.78^{+0.04}_{-0.04}$, and the inclination angle of the inner disk $i=29.2^{+0.3}_{-0.5}$ degrees. \citet{kumarEstimationSpinMass2022} analyzed nearly simultaneous data from the Neutron Star Interior Composition Explorer (\textit{NICER}, 0.6-10.0 keV, \citealt{gendreauNeutronStarInterior2012}) and \textit{NuSTAR} (3.0-79.0 keV) during the soft and hard states and the spin $a_{*}=0.80^{+0.02}_{-0.02}$ and the mass of the black hole $M=8.7^{+0.3}_{-0.3}M_{\odot}$ were estimated from modeling the soft state spectrum of the source by using the CF method. The inclination of the source was determined to be $i=(36.5\pm1.0)^{\circ}$. 

In this paper, we have performed an analysis of the Hard X-ray Modulation Telescope (\textit{Insight-HXMT}, \citealt{zhangIntroductionHardXray2014}) data on MAXI J1348-630 at IMS. Using the relativistic reflection model $\mathtt{relxill}$ family \citep{garciaImprovedReflectionModels2014a}, we obtained the constraints on the spin of the black hole and the inclination angle of the accretion disk. We also inferred the ionization state and iron abundance by X-ray reflection fitting.

The paper is organized as follows. The description of the observed data and the data reduction are described in Section \ref{sec2}. The results of the spectral modeling analysis are described in Section \ref{sec3}. A discussion is given in Section \ref{sec4}, and the conclusion is summarized in Section \ref{sec5}.

\section{Observations and Data Reduction} \label{sec2}
\begin{figure}
 \includegraphics[width=\columnwidth]{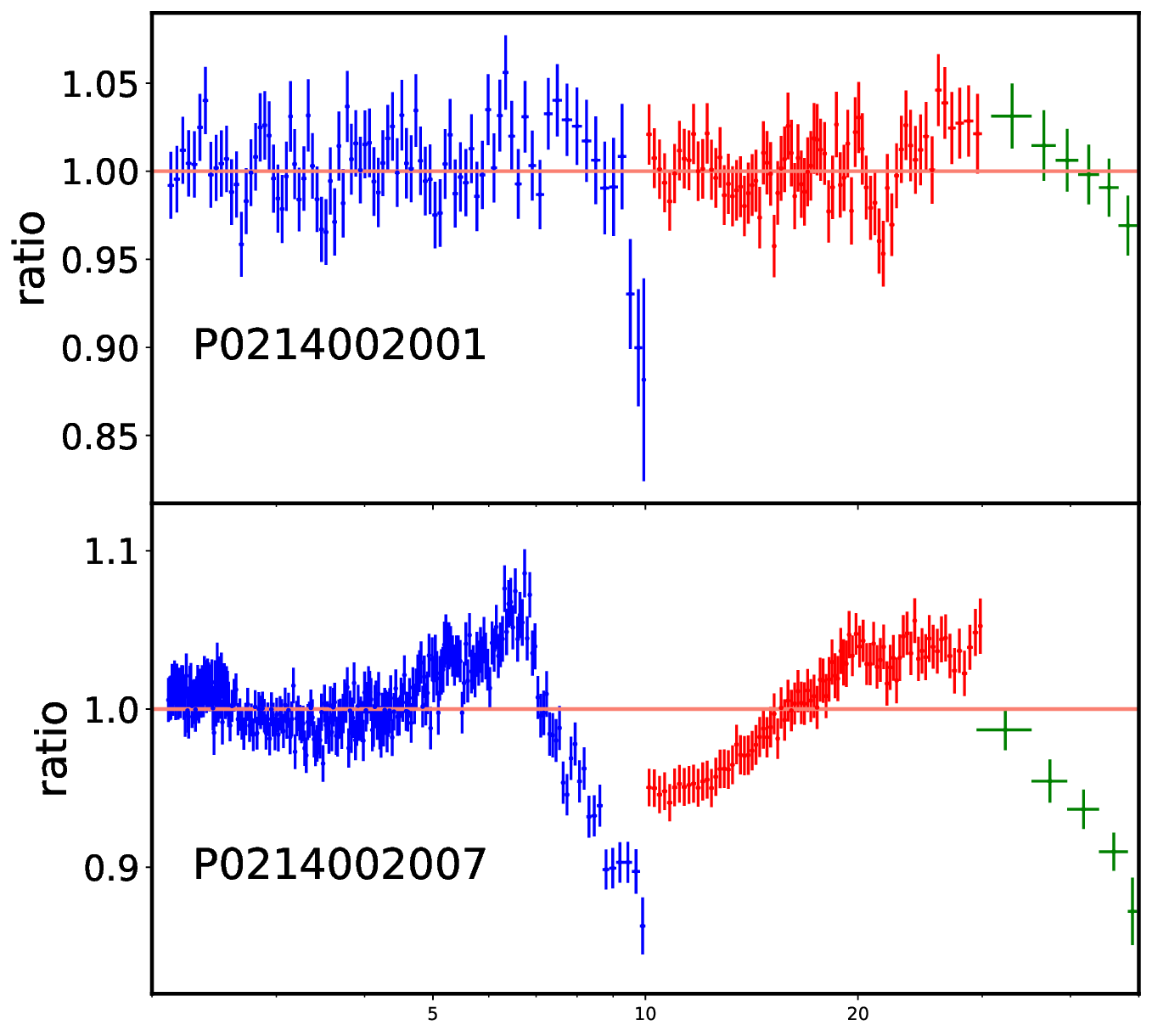}
 \caption{Fits for different observation: The data were fitted over 2.1–50.0 keV, ignoring 6.0–8.0 keV and 20.0–40.0 keV. LE, ME and HE data are plotted in blue, red and green, respectively. The data have been rebinned for display clarity. The observation number is shown in the bottom left corner of each panel. Upper panel: No significant features. Lower panel: Relativistic reflection features shown as a broadened iron line and Compton hump.}
 \label{fig:com}
\end{figure}

\begin{figure}
 \includegraphics[width=\columnwidth]{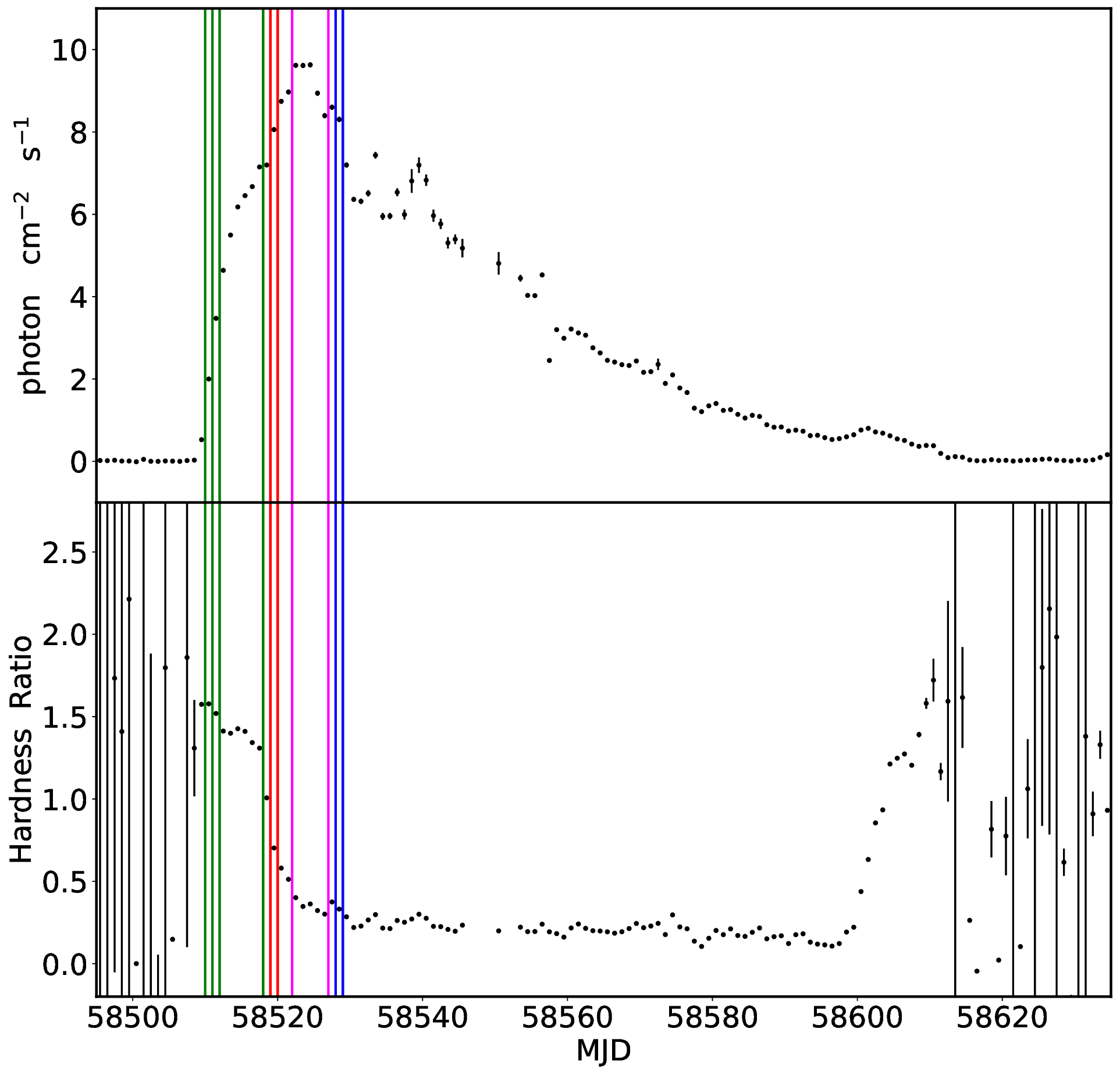}
 \caption{Upper panel: \textit{MAXI}/GSC observations of MAXI J1348-630 in 2.0-20.0 keV. The green, red, magenta and blue vertical lines represent the LHS, HIMS, SIMS and HSS observations of \textit{Insight-HXMT}, respectively. Lower panel: time evolution of the hardness ratio (4-20 keV/2-4 keV).}
 \label{fig:HR}
\end{figure}

\textit{Insight-HXMT}, China's first X-ray astronomical satellite, can not only perform wide-band, large-field X-ray sky survey, but also study black holes, neutron stars and other high-energy celestial bodies with short timescale variability and wide-band energy spectrum. Meanwhile, it is a highly sensitive all-day gamma-ray burst monitor. It was successfully launched on June 15, 2017. MAXI J1348-630 was observed regularly with \textit{Insight-HXMT} from 2019 January 27 to July 29. There were two significant outbursts during this period: the first outburst lasted four months and the source showed all four typical spectral states. And the second one started in May. We select the observations from the beginning of the first outburst for spin analysis in this work. In this period, all of the observations enabling access to spectra products for analysis have been applied to preliminary data analysis, as described in the Sec~\ref{pre}.  The exposure time and the flux for each spectra are listed in Table~\ref{tab:1}. The unabsorbed  0.001–100 keV disk luminosity was obtained by the flux. Taking the BH mass at $8M_{\odot}$ and a distance of 2.2kpc \citep{chauhanMeasuringDistanceBlack2021}. We also calculated the Eddington-scaled disk luminosity $\mathrm{L(L_{edd})}$ and $\mathrm{L_{edd}}=1.3\times10^{38}(M/M_{\odot})$. According to \citet{dunnGlobalStudyBehaviour2011}, most of the transition from the power-law-dominated to disc-dominated states (disk fraction>80\%) occurs between 10\% and 30\% $\mathrm{L_{Edd}}$. The results we have obtained are consistent with this. However, considering that the distances we used are at the lower end of the possible range, it is possible to obtain a higher luminosity. For convenience, we will refer to use SP1-SP10 as shorthand for the 10 spectra. There are some spectra at the beginning of the observations where the reflection features are not obvious, see Figure~\ref{fig:com}. So finally 6 good spectra (SP5-SP10) with clearly reflective features were selected for the spin measurement. Of these, SP5-SP6 is in HIMS; SP7-SP8 is in SIMS and SP9-SP10 is in HSS  \citep{zhangPeculiarDiskBehaviors2022}. We obtained the daily averaged light curve from \textit{MAXI}/GSC \citep{matsuokaMAXIMissionISS2009}, and all of the selected \textit{Insight-HXMT} observation dates are marked with vertical lines in Figure~\ref{fig:HR}. 

\begin{figure}
\centering
\includegraphics[scale=0.35,angle=270]{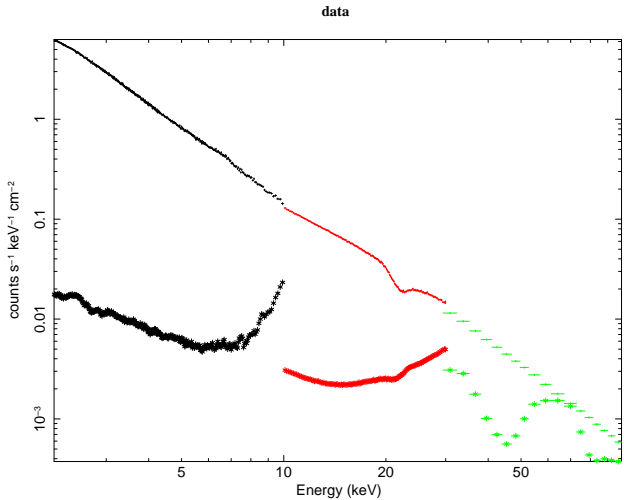}
\caption{The data and background counts of SP5 are shown with the logarithmic y-axis. When above 50 keV, the accounts of background photons become close to data.}
 \label{fig:back}
\end{figure}

We followed the recommendations given by the \textit{Insight-HXMT} team and the standard procedures for individual instruments. Data pipelines and tools using \textit{Insight-HXMT} Data Analysis Software (HXMTDAS) v2.05. The \textit{Insight-HXMT} spectra are extracted from cleaned event files filtered by good time intervals (GTIs). The GTIs recommended by pipeline are intervals when (1) elevation angle greater than 10 degrees; (2) geomagnetic cut-off rigidities greater than 8 GeV; (3) satellite not in SAA and spaced 300 seconds near SAA; (4) pointing deviation to the source less than 0.04 degrees. Multiple short exposures on the same day are merged with $\mathtt{addspec}$ to increase the signal-to-noise ratio. All data were grouped to achieve at least 100 photons per energy bin using the FTOOLS command $\mathtt{grppha}$. Then, we add a $1\%,2\%,2\%$  system uncertainty to LE, ME and HE spectra, respectively, to account for the instrument uncertainty. For spectral analysis, we use data in energy bands of 2.1-10 keV, 10-30 keV, and 30-50 keV for LE, ME, and HE, respectively. The background dominates the spectra at higher energies (Figure~\ref{fig:back}).

\begin{table*}
	\centering
	\fontsize{8}{12}\selectfont    
	\caption{Observation log of MAXI J1348-630}
        \label{tab:1}
	\begin{tabular}{cccccccc}
		\toprule
		ObsID & Start Time&\ End Time  & Flux &$\mathrm{L}$ & Spectra & State \cr
                &    MJD  & MJD & $\mathrm{\times 10^{-7}(erg\ cm^{-2}\ s^{-1})}$& $\mathrm{(L_{edd})}$& & &\\
            
            \midrule
            P021400200101 & 58510.253 & 58510.465 & 0.5264  & 0.031 & SP1 &LHS
            \\
            \midrule
            P021400200201 & 58511.313 & 58511.524 & 0.9653  & 0.056 & SP2 & LHS
            \\
            \midrule
            P021400200301 & 58512.308 & 58512.520 & 1.2639  & 0.074 & SP3 & LHS
            \\
            \midrule
            P021400200601 & 58518.011 & 58518.150 & \multirow{3}{*}{1.7515}&\multirow{3}{*}{0.102}& \multirow{3}{*}{SP4} & \multirow{3}{*}{LHS}\cr
            P021400200602 & 58518.150 & 58518.294 & & & &\cr
            P021400200603 & 58518.294 & 58518.423 & & & &\\
            \midrule
            P021400200701 & 58519.007 & 58519.145 & \multirow{4}{*}{1.7857}  &\multirow{4}{*}{0.104}& \multirow{4}{*}{SP5} &\multirow{4}{*}{HIMS} \cr
            P021400200702 & 58519.145 & 58519.287 & & & & \cr
            P021400200703 & 58519.287 & 58519.426 & & & & \cr
            P021400200704 & 58519.426 & 58519.617 & & & & \\
            \midrule
            P021400200801 & 58519.803 & 58519.932 & \multirow{4}{*}{1.8808}& \multirow{4}{*}{0.110}& \multirow{4}{*}{SP6}&\multirow{4}{*}{HIMS}  &\cr
            P021400200802 & 58519.932 & 58520.067 & & & &\cr
            P021400200803 & 58520.067 & 58520.212 & & & &\cr
            P021400200804 & 58520.212 & 58520.414 & & & &\\
            \midrule
            P021400201101 & 58522.591 & 58522.757 & \multirow{3}{*}{2.3724}&\multirow{3}{*}{0.139}& \multirow{3}{*}{SP7} &\multirow{3}{*}{SIMS}\cr
            P021400201102 & 58522.757 & 58522.890 & & & &\cr
            P021400201103 & 58522.890 & 58523.069 & & & & \\
            \midrule
            P021400201701 & 58527.298 & 58527.461 &  \multirow{4}{*}{2.1462}&\multirow{4}{*}{0.125}& \multirow{4}{*}{SP8}&\multirow{4}{*}{SIMS}  \cr
            P021400201702 & 58527.461 & 58527.593 & & & &\cr
            P021400201703 & 58527.593 & 58527.726 & & & &\cr
            P021400201704 & 58527.725 & 58527.904 & & & &\\
            \midrule
            P021400201801 & 58528.624 & 58528.786 &\multirow{3}{*}{2.0973} &\multirow{3}{*}{0.123}& \multirow{3}{*}{SP9} &\multirow{3}{*}{HSS} \cr
            P021400201802 & 58528.786 & 58528.918 & & & &\cr
            P021400201803 & 58528.918 & 58529.051 & & & &\\
            \midrule
            P021400201804 & 58529.050 & 58529.183  &\multirow{3}{*}{2.0393}&\multirow{3}{*}{0.119}& \multirow{3}{*}{SP10} &\multirow{3}{*}{HSS} \cr
            P021400201805 & 58529.183 & 58529.316 & & & &\cr
            P021400201806 & 58529.316 & 58529.423 & & & &\\
   \bottomrule
	\end{tabular}\vspace{0cm}
        \begin{tablenotes} 
		\item Notes: Flux is calculated in 0.001-100 keV  with the $\mathtt{cflux}$ command in XSPEC. For the calculation of the Eddington luminosity, we take the BH mass at 8 $M_{\odot}$, the distance of 2.2 kpc. 
     \end{tablenotes} 
	\label{tab:Training_sizes}
\end{table*}

\section{Spectral Analysis and Results}\label{sec3}
The spectral analysis is performed using XSPEC version 12.12.1 \citep{arnaudXSPECFirstTen1996}, which is included as part of HEASOFT v6.30. All models include a multiplicative constant component to account for differences in flux calibration between instruments. There is no overlap in the coverage among the 3 instruments of \textit{Insight-HXMT}. This constant is fixed at 1.0 for LE and allowed to vary for ME($C_{1}$) and HE($C_{2}$). Unless otherwise noted, all uncertainties reported in this paper are given at the $90\%$ confidence level. We use the $\mathtt{tbabs}$ component to model the neutral Galactic absorption, using the abundances of \citet{wilmsAbsorptionRaysInterstellar2000} and the cross sections of \citet{vernerAtomicDataAstrophysics1996}. 

\subsection{Preliminary Spectral Analysis}\label{pre}

First, we use the simple absorption power-law model $\mathtt{const*tbabs (diskbb+powerlaw)}$ (Model 1) to fit all ten spectra. Since the column density cannot be constrained, we fix the column density ($N_{\mathrm{H}}$) at $0.6\times10^{22} \mathrm{cm^{-2}}$, which is the fitting result from \citet{mallBroadbandSpectralProperties2023c} using AstroSat spectroscopy. When we fit spectra at this point, we ignore $6.0-8.0\ \mathrm{keV},20.0-40.0\ \mathrm{keV}$ to demonstrate the existence of reflection components. According to \citet{tominagaDiscoveryBlackHole2020}, it can be known that the spectrum is in the intermediate state. For example, from the residual profiles Figure~\ref{fig:com}, we can clearly see the broadened iron profile of 6-10 keV and the Compton hump feature above 15 keV in the lower panel. Finally, we filtered out six spectra (SP5-SP10) with distinctive reflection features for subsequent fitting analysis.

\begin{table}
	\centering
	\fontsize{8}{12}\selectfont    
	\caption{The changes in $\chi^{2}_{\nu}$ for Model 1 and Model 2. The equivalent width of the Gaussian iron is also shown.  }
        \label{tab:gauss}
	\begin{tabular}{ccccc}
		\toprule
		\multirow{2}{*}{Spectra}&$\chi^{2}$&$\chi^{2}$& \multirow{2}{*}{$\nu$} &EW
            \cr
		  & model 1 & model 2 & & (keV) 
		  \\
            \midrule
            LHS &&&&
            \\
            SP1 & 1097.82(0.85) & 1092.64(0.85) & 1285 & 0.01 \cr
            SP2 & 1303.73(1.01) & 1278.54(0.99) & 1285 & 0.02 \cr
            SP3 & 1341.51(1.04) & 1297.89(1.01) & 1285 & 0.10 \cr
            SP4 & 1667.33(1.30) & 1307.55(1.02) & 1285 & 0.20 \\
            \midrule
            HIMS &&&&
            \\
            SP5 & 2585.93(2.01) & 1749.36(1.36) & 1285 & 0.42 \cr
            SP6 & 1974.43(1.56) & 1397.12(1.10) & 1267 & 0.31 \\
            \midrule
            SIMS &&&&
            \\
            SP7 & 2251.70(1.80) & 1297.89(0.99) & 1248 & 0.39 \cr
            SP8 & 5171.45(4.02) & 2850.84(2.22) & 1285 & 0.47 \\
            \midrule
            HSS &&&&
            \\
            SP9 & 5975.02(4.65) & 2834.29(2.21) & 1285 & 0.51 \cr
            SP10 & 3058.65(2.38) & 1928.22(1.50) & 1285 & 0.57 \\
            \midrule
    \end{tabular}
\end{table}

We then use a preliminary phenomenological model $\mathtt{const*tbabs*(diskbb+gaussian+cutoffpl)}$ (Model 2) to fit the spectrum. The model $\mathtt{gaussian}$ represents the iron $\mathrm{K}\alpha$ emission line. The central energy of the Gaussian line profile is limited to 6.4-6.97 keV. With the new model configuration the fit is effectively improved, see Table~\ref{tab:gauss}. e.g. SP5 has a reduced from $\chi^{2}_{\nu} =2.01$ to $\chi^{2}_{\nu} =1.36$. Although $\mathtt{diskbb}$ and $\mathtt{gaussian}$ fit the disk and emission line components well respectively, there are still Compton humps above 15 keV and a high energy cutoff in the data model ratio plot. 

\subsection{Relativistic Reflection Model}
\subsubsection{Incident Spectrum with A High-energy Cutoff Power-Law}

We then replace the power-law model with the reflection model $\mathtt{relxill (relxill\ v2.0)}$\footnote{\url{http://www.sternwarte.uni-erlangen.de/~dauser/research/relxill/}} \citep{garciaXRAYREFLECTEDSPECTRA2010} to account for the relativistically fuzzy reflection components in the data. The complete model is $\mathtt{const*tbabs(diskbb+relxill)}$ (M
Model 3) . As mentioned above, we set the initial value of $N_{\mathrm{H}}$ at $0.6\times10^{22}\mathrm{cm^{-2}}$ to allow free fitting. We found that this model can provide a good constraint of $N_{\mathrm{H}}$ and the fitting value was very close to it, so we set $N_{\mathrm{H}}$ all at $0.6\times10^{22}\mathrm{cm^{-2}}$ for fitting analysis. Assuming $R_{\mathrm{in}}=R_{\mathrm{ISCO}}$, the $R_{\mathrm{in}}$ was fixed at -1. We tried to free both emissivity index $q$  and break radius $R_{br}$ to fit, but this could not give an upper bound for $q$. Because the emissivity index $q$ of the reflection emissivity profile $\epsilon(r)\propto r^{-q}$ and the break radius $R_{br}$ cannot be constrained at the same time. So we assume canonical case: $q_{\mathrm{in}}=q_{\mathrm{out}}=3$ \citep[e.g.][]{fabianXrayFluorescenceInner1989}. We set the cutoff energy at a typical and reasonable value of $E_{\mathrm{cut}}=300$ keV following the traditional case \citep{bardeenRotatingBlackHoles1972a,rossComprehensiveRangeXray2005}. The outer radius of disk ($R_{\mathrm{out}}$) was fixed at default value of 400 $R_{\mathrm{g}}$ ($R_{\mathrm{g}}=\mathrm{GM}/c^{2}$). In addition, MAXI J1348-630 is a galactic source thus the redshift ($z$) was set to zero. All other parameters are allowed to vary freely including the inclination angle ($i$), photon index ($\Gamma$), ionization state of the accretion disk ($\mathrm{log}\xi$), iron abundance ($A_{\mathrm{Fe}}$), reflection fraction ($R_{\mathrm{ref}}$), and normalization of $\mathtt{relxill}$ ($N_{\mathrm{relxill}}$). After considering the relativistic reflection component, our model gives much better fitting results, which are listed in detail in Table~\ref{tab:2}.

\begin{table*}
	\centering
	\fontsize{8}{12}\selectfont    
	\caption{Best-fitting parameters with Model 3, inner radius were frozen at $R_{\mathrm{ISCO}}$}
        \label{tab:2}
	\begin{tabular}{cccccccc}
		\toprule
		\multirow{3}{*}{\shortstack{Spectral\\ Component}}&\multirow{2}{*}{Parameter}&\multicolumn{2}{c}{HIMS} &\multicolumn{2}{c}{SIMS} &\multicolumn{2}{c}{HSS}  \cr
		\cmidrule(lr){3 -4}  \cmidrule(lr){5 -6} \cmidrule(lr){7-8} 
		 && SP5 & SP6 & SP7 & SP8 & SP9 & SP10  \cr
        \midrule
        Tbabs & $N_{\mathrm{H}}(\times10^{22}\mathrm{cm}^{-2})$ & $0.6^{\dagger}$ & $0.6^{\dagger}$ & $0.6^{\dagger}$ & $0.6^{\dagger}$ & $0.6^{\dagger}$ & $0.6^{\dagger}$ 
        \\
        \midrule
	\multirow{2}{*}{Diskbb}
        & $kT_{\mathrm{in}}(\mathrm{keV})$ & $0.608^{+0.013}_{-0.008}$ & $0.640^{+0.003}_{-0.007}$ & $0.776^{+0.005}_{-0.004}$ & $0.752^{+0.002}_{-0.002}$ & $0.736^{+0.002}_{-0.001}$ & $0.732^{+0.002}_{-0.002}$  \cr
        & $\mathrm{N_{Diskbb}}(\times10^{4})$ & $2.08^{+0.16}_{-0.20}$ & $2.07^{+0.16}_{-0.09}$ & $2.16^{+0.04}_{-0.07}$ & $2.20^{+0.02}_{-0.02}$ & $2.69^{+0.02}_{-0.02}$ & $2.89^{+0.04}_{-0.04}$ 
        \\
        \midrule
        \multirow{7}{*}{relxill}
        & $a_{*}$ & 
        $>0.85$ & 
        $>0.79$ & 
        $0.85^{+0.13}_{-0.34}$ & $0.80^{+0.07}_{-0.09}$ & $0.76^{+0.09}_{-0.12}$ & $0.80^{+0.12}_{-0.15}$  \cr
        & $i(\mathrm{deg})$ & $35.5^{+2.7}_{-2.5}$ & $33.2^{+4.1}_{-6.5}$ & $33.1^{+2.9}_{-3.2}$ & $27.3^{+2.0}_{-2.4}$ & $26.4^{+2.0}_{-2.8}$ & $26.5^{+2.9}_{-4.2}$  \cr
        & $\Gamma$ & 
        $2.15^{+0.02}_{-0.02}$ & $2.16^{+0.02}_{-0.04}$ & $2.39^{+0.02}_{-0.04}$ & $2.35^{+0.01}_{-0.01}$ & $2.36^{+0.01}_{-0.01}$ & $2.30^{+0.01}_{-0.02}$  \cr
        & $\mathrm{log}\xi$ & $3.83^{+0.15}_{-0.22}$ & $4.17^{+0.32}_{-0.20}$ & $4.57^{+0.13}_{-0.17}$ & $4.37^{+0.07}_{-0.08}$ & $4.31^{+0.08}_{-0.05}$ & $4.21^{+0.09}_{-0.12}$  \cr
        & $A_{\mathrm{Fe}}$ & $3.8^{+1.5}_{-1.1}$ & 
        $>4.0$ & 
        $>5.2$ & 
        $>9.5$ & 
        $>9.5$ & 
        $>9.5$  \cr
        & $R_{\mathrm{ref}}$ & $0.56^{+0.20}_{-0.12}$ & $0.62^{+0.17}_{-0.17}$ & $0.57^{+0.19}_{-0.12}$ & $0.52^{+0.06}_{-0.06}$ & $0.64^{+0.08}_{-0.06}$ & $1.05^{+0.15}_{-0.18}$  \cr
        & $\mathrm{norm}$ & $0.21^{+0.02}_{-0.02}$ & $0.19^{+0.03}_{-0.04}$ & $0.18^{+0.03}_{-0.02}$ & $0.15^{+0.01}_{-0.01}$ & $0.09^{+0.004}_{-0.007}$ & $0.06^{+0.003}_{-0.007}$ 
        \\
        \midrule
        \multirow{2}{*}{constant}
        & $C_{1}$ & 1.00 & 1.02 & 1.00 & 1.02 & 0.98 & 1.03  \cr
        & $C_{2}$ & 0.95 & 1.00 & 1.02 & 1.00 & 1.02 & 1.02   
        \\
        \midrule
        $\chi^{2}/\nu$ & & 903.61/1285 & 938.42/1267 & 926.94/1248 & 1466.98/1285 & 1584.74/1285 & 1543.48/1285 
        \\
        \midrule
        $\chi^{2}_{\nu}$   & & 0.70 & 0.74 & 0.74 & 1.14 & 1.23 & 1.20
        \\
        
	\bottomrule
	\end{tabular}\vspace{0cm}
        \begin{tablenotes} 
		\item Notes: The best-fitting parameters obtained by \textit{Insight-HXMT} observations with model $\mathtt{constant*tbabs*(diskbb+relxill)}$. The parameters with the symbol “$\dagger$” indicate they are fixed at values given. 
     \end{tablenotes} 
	\label{tab:Training_sizes}
    \end{table*}

\subsubsection{Incident Spectrum with The Comptonization Continuum}

\begin{figure}
\centering
\includegraphics[width=\columnwidth]{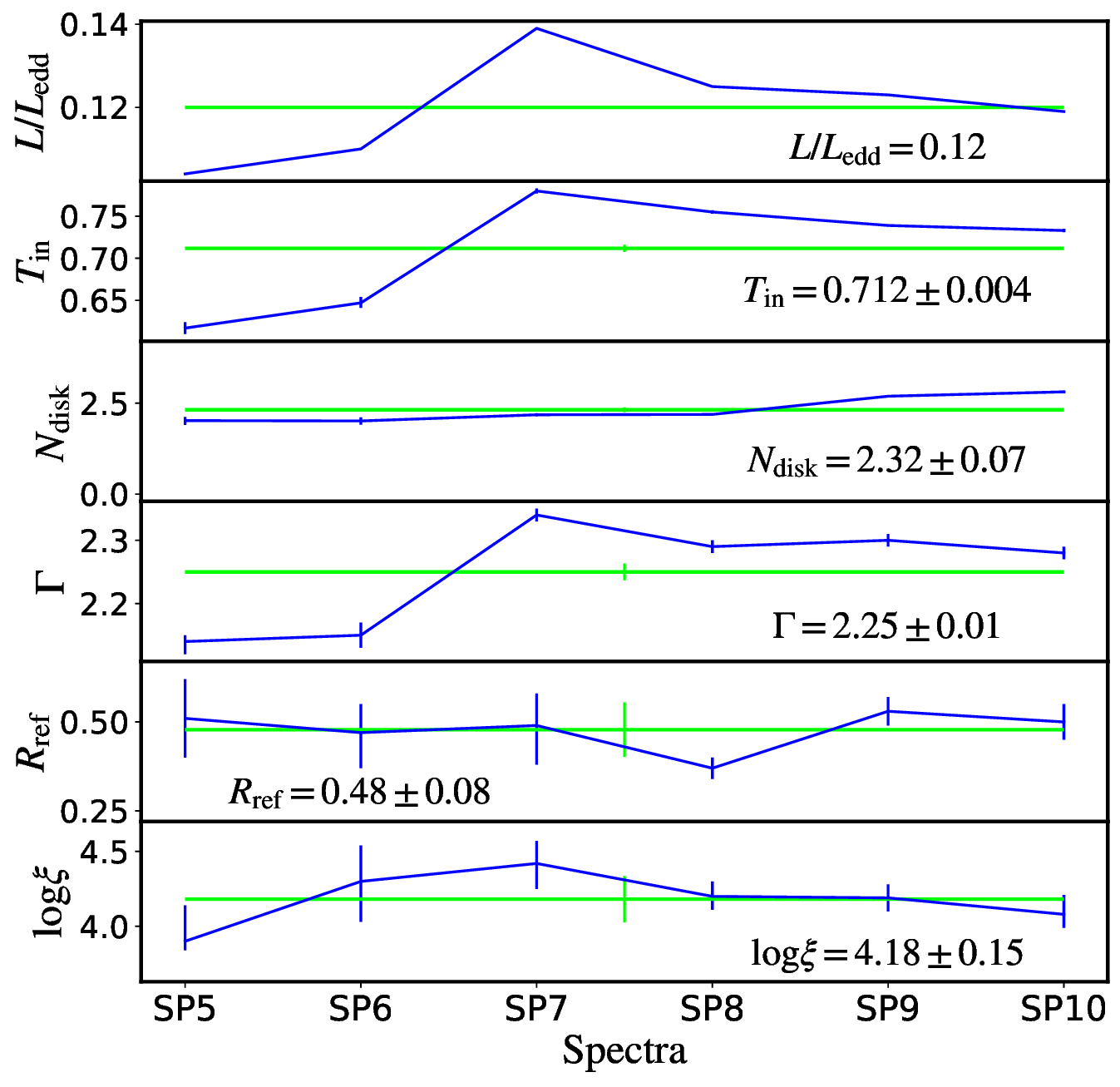}
\caption{The fitted parameters versus the spectra ID with Model 4. $L/L_{\mathrm{edd}}$ is the luminosity. $T_{\mathrm{in}}$ is the temperature at the inner disk radius (keV). $N_{\mathrm{disk}}$ is the normalization of the $\mathtt{diskbb}$ model ($N_{\mathrm{disk}}=(R_{\mathrm{in}}/D_{10})^{2}\times \mathrm{cos\theta}$). $\Gamma$ is the power law index of the primary source spectrum in $\mathtt{relxillCp}$ model.  $R_{\mathrm{ref}}$ is the reflection fraction parameter. $\mathrm{log\xi}$ is the ionization of the accretion disk. The average results are shown in black font.}
 \label{fig:table3}
\end{figure}

\begin{figure}
\centering
\includegraphics[width=\columnwidth]{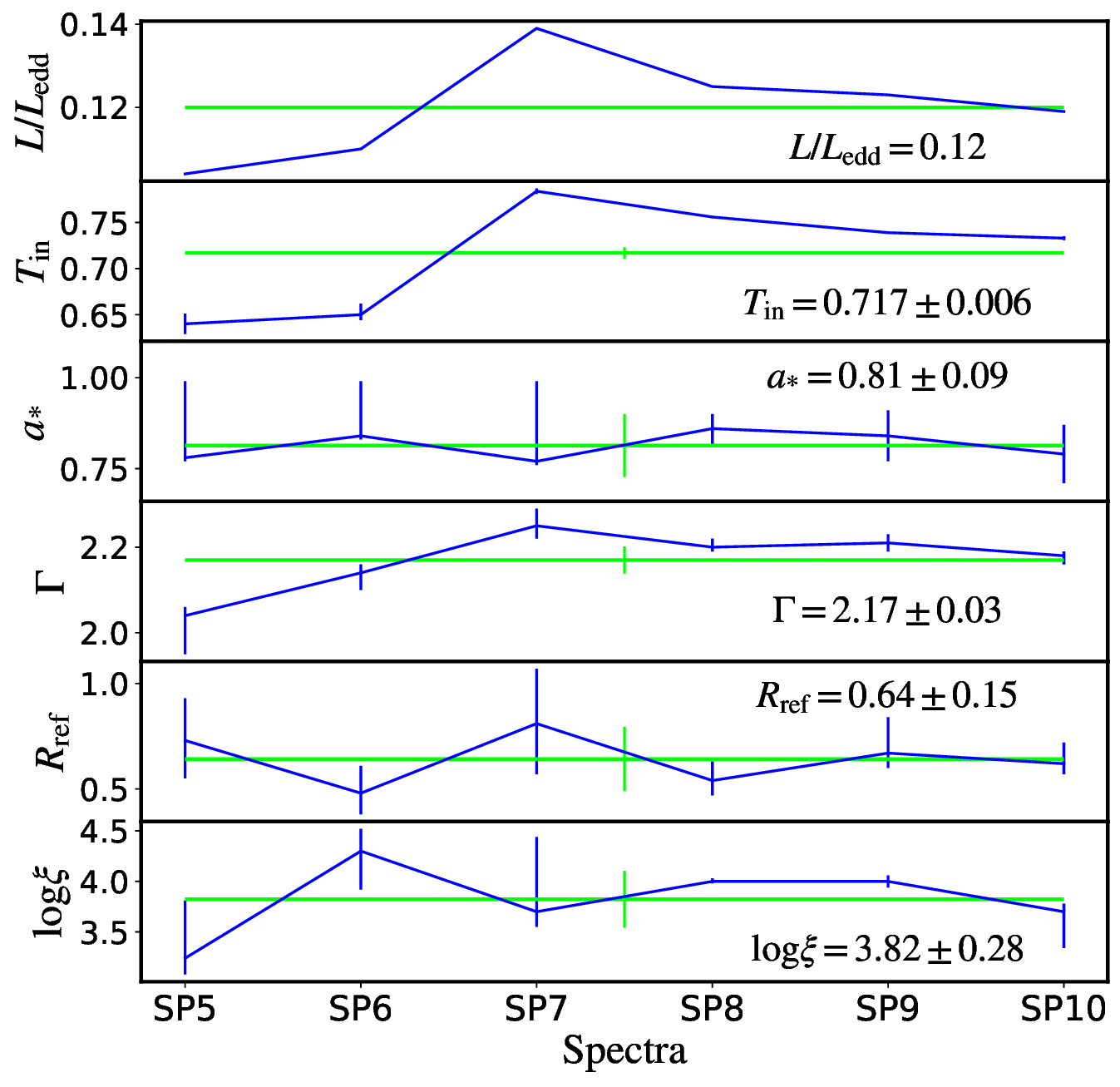}
\caption{The fitted parameters versus the spectra ID with Model 5. $L/L_{\mathrm{edd}}$ is the luminosity. $T_{\mathrm{in}}$ is the temperature at the inner disk radius (keV). $a_{\mathrm{*}}$ is the spin of the black hole in dimensionless units. $\Gamma$ is the power law index of the primary source spectrum in $\mathtt{relxillCp}$ model. $R_{\mathrm{ref}}$ is the reflection fraction parameter. $\mathrm{log\xi}$ is the ionization of the accretion disk. The average results are shown in black font.}
 \label{fig:table4}
\end{figure}

\begin{figure*}
 \includegraphics[scale=0.7,angle=270]{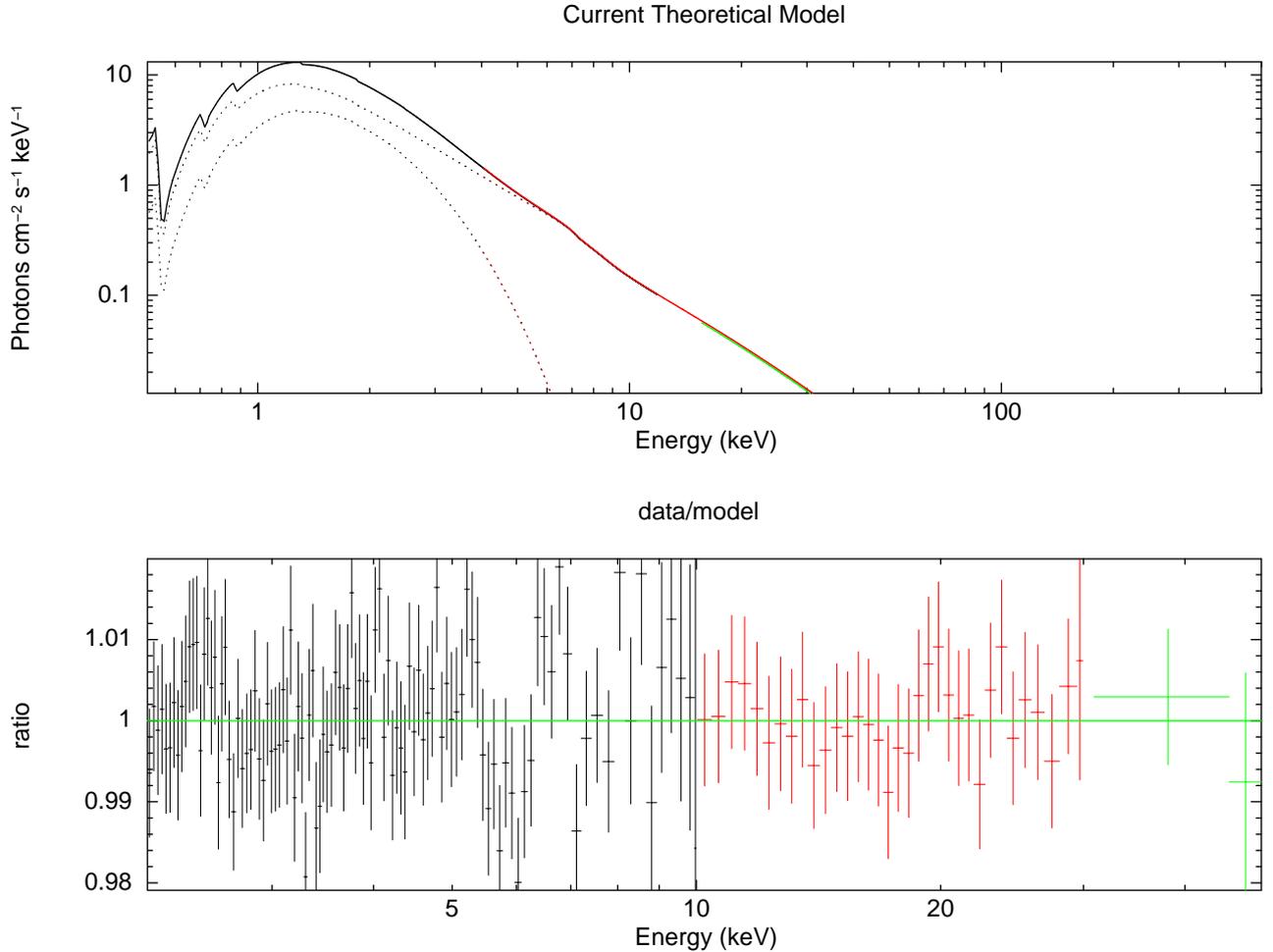}
 \caption{Upper panel: The best-fitting model $\mathtt{const*tbabs(diskbb+relxillCp)}$ (Model 5) for SP5. Bottom panel: The ratio of fitting Insight-HXMT observations (model to data). LE, ME and HE data are plotted in black, red and green, respectively. The data have been rebinned for display clarity.}
 \label{fig:7cp}
\end{figure*}

Next, we replaced $\mathtt{relxill}$ with $\mathtt{relxillCp}$, which contains a more physical Compton continuum case. The emissivity of both models can be modeled by an empirical broken power law. The difference between them is that the primary spectrum of the $\mathtt{relxill}$ is $\mathtt{cutoffpl}$, while another is a $\mathtt{nthcomp}$ primary spectrum and the density is a free parameter in the model \citep{zdziarskiBroadbandRayXray1996,zycki1989MayOutburst1999}. So the latter model is closer to the real case. The complete model is $\mathtt{const*tbabs(diskbb+relxillCp)}$. To verify the reliability of the hypothesis of  $R_{\mathrm{in}}=R_{\mathrm{ISCO}}$, we first free the $R_{\mathrm{ISCO}}$ and fixed $a_{*}$ at maximum 0.998  (Model 4) . In this model, $\mathrm{log}N$ is the logarithmic value of the disk density (in $\mathrm{cm}^{−3}$). For consistency with the above, the $\mathrm{log}N$ was fixed at 15 which is the same as the model $\mathtt{relxill}$. And the electron temperature in the corona ($kT_{e}$) was fixed at 100 keV based on the relationship between $kT_{e}$ and $E_{\mathrm{cut}}$: $kT_{e}=1/3E_{\mathrm{cut}}$ \citep{petrucciTestingComptonizationModels2001a}. The fitting results are listed in Table~\ref{tab:3} and Figure~\ref{fig:table3}. Then we set the density of disk ($\mathrm{log}N$) free to spin analysis (Model 5), the results of all fitted parameters in Model 5 are shown in Table~\ref{tab:4} and Figure~\ref{fig:table4}, and the SP5 is shown in Figure~\ref{fig:7cp} as a representative spectral fitting result.

\begin{table*}
	\centering
	\fontsize{8}{12}\selectfont    
	\caption{Best-fitting parameters with Model 4, $a_{*}$ is fixed at 0.998}
        \label{tab:3}
	\begin{tabular}{cccccccccc}
		\toprule
		\multirow{3}{*}{\shortstack{Spectral\\ Component}}&\multirow{2}{*}{Parameter}&\multicolumn{2}{c}{HIMS} &\multicolumn{2}{c}{SIMS} &\multicolumn{2}{c}{HSS}  \cr
		\cmidrule(lr){3 -4}  \cmidrule(lr){5 -6} \cmidrule(lr){7-8} 
		 && SP5 & SP6 & SP7 & SP8 & SP9 & SP10  \cr
        \midrule
        Tbabs & $N_{\mathrm{H}}(\times10^{22}\mathrm{cm}^{-2})$ & $0.6^{\dagger}$ & $0.6^{\dagger}$ & $0.6^{\dagger}$ & $0.6^{\dagger}$ & $0.6^{\dagger}$ & $0.6^{\dagger}$ 
        \\
        \midrule
	\multirow{2}{*}{Diskbb}
        & $kT_{\mathrm{in}}(\mathrm{keV})$ & $0.617^{+0.007}_{-0.007}$ & $0.647^{+0.007}_{-0.006}$ & $0.780^{+0.003}_{-0.003}$ & $0.755^{+0.002}_{-0.002}$ & $0.739^{+0.001}_{-0.001}$ & $0.733^{+0.002}_{-0.002}$  \cr
        & $\mathrm{N_{Diskbb}}(\times10^{4})$ & $2.02^{+0.10}_{-0.12}$ & $2.01^{+0.10}_{-0.10}$ & $2.18^{+0.04}_{-0.04}$ & $2.19^{+0.02}_{-0.02}$ & $2.69^{+0.02}_{-0.02}$ & $2.81^{+0.03}_{-0.04}$ 
        \\
        \midrule
        \multirow{7}{*}{relxillCp}
        & $R_{\mathrm{in}}(R_{\mathrm{ISCO}})$ & 
        $<1.84$ & 
        $<2.01$ & 
        $1.94^{+0.39}_{-0.40}$ & $2.51^{+0.52}_{-0.27}$ & $2.50^{+0.31}_{-0.25}$ & $3.08^{+0.76}_{-0.56}$  \cr
        & $i(\mathrm{deg})$ & $35.3^{+2.5}_{-2.3}$ & $33.4^{+4.9}_{-4.2}$ & $32.5^{+2.2}_{-3.1}$ & $26.2^{+2.2}_{-3.8}$ & $26.6^{+1.9}_{-2.3}$ & $22.6^{+4.5}_{-4.8}$  \cr
        & $\Gamma$ & 
        $2.14^{+0.01}_{-0.02}$ & $2.15^{+0.02}_{-0.02}$ & $2.34^{+0.01}_{-0.01}$ & $2.29^{+0.01}_{-0.01}$ & $2.30^{+0.01}_{-0.01}$ & $2.28^{+0.01}_{-0.01}$  \cr
        & $\mathrm{log}\xi$ & $3.90^{+0.24}_{-0.06}$ & $4.30^{+0.24}_{-0.27}$ & $4.42^{+0.15}_{-0.17}$ & $4.20^{+0.10}_{-0.09}$ & $4.19^{+0.09}_{-0.09}$ & $4.08^{+0.13}_{-0.09}$  \cr
        & $A_{\mathrm{Fe}}$ & $4.3^{+0.8}_{-1.1}$ & 
        $>4.5$ & 
        $>7.9$ & 
        $>9.9$ & 
        $>9.9$ & 
        $>9.8$  \cr
        & $R_{\mathrm{ref}}$ & $0.51^{+0.11}_{-0.11}$ & $0.47^{+0.08}_{-0.10}$ & $0.49^{+0.09}_{-0.11}$ & $0.37^{+0.03}_{-0.03}$ & $0.53^{+0.04}_{-0.04}$ & $0.50^{+0.05}_{-0.05}$  \cr
        & $\mathrm{norm}$ & $0.20^{+0.02}_{-0.02}$ & $0.19^{+0.02}_{-0.02}$ & $0.15^{+0.01}_{-0.01}$ & $0.13^{+0.004}_{-0.004}$ & $0.08^{+0.003}_{-0.003}$ & $0.06^{+0.003}_{-0.004}$ &  \\
        \midrule
        \multirow{2}{*}{constant}
        & $C_{1}$ & 1.00 & 1.02 & 0.99 & 0.98 & 0.95 & 0.97  \cr
        & $C_{2}$ & 0.97 & 1.01 & 1.04 & 0.98 & 1.01 & 0.95   
        \\
        \midrule
        $\chi^{2}/\nu$ & & 906.93/1285 & 940.47/1267 & 942.19/1248 & 1714.91/1285 & 1719.99/1285 & 1626.16/1285 
        \\
        \midrule
        $\chi^{2}_{\nu}$   & & 0.70 & 0.74 & 0.75 & 1.23 & 1.28 & 1.22 
        \\
        
	\bottomrule
	\end{tabular}\vspace{0cm}
        \begin{tablenotes} 
		\item Notes: The best-fitting parameters obtained by \textit{Insight-HXMT} observations with model $\mathtt{constant*tbabs*(diskbb+relxillCp)}$. The parameters with the symbol “$\dagger$” indicate they are fixed at values given. 
     \end{tablenotes} 
\end{table*}

\begin{table*}
	\centering
	\fontsize{8}{12}\selectfont    
	\caption{Best-fitting parameters with Model 5, inner radius is fixed at $R_{\mathrm{ISCO}}$}
        \label{tab:4}
	\begin{tabular}{cccccccccc}
		\toprule
		\multirow{3}{*}{\shortstack{Spectral\\ Component}}&\multirow{2}{*}{Parameter}&\multicolumn{2}{c}{HIMS} &\multicolumn{2}{c}{SIMS} &\multicolumn{2}{c}{HSS}  \cr
		\cmidrule(lr){3 -4}  \cmidrule(lr){5 -6} \cmidrule(lr){7-8} 
		 && SP5 & SP6 & SP7 & SP8 & SP9 & SP10  \cr
        \midrule
        Tbabs & $N_{\mathrm{H}}(\times10^{22}\mathrm{cm}^{-2})$ & $0.6^{\dagger}$ & $0.6^{\dagger}$ & $0.6^{\dagger}$ & $0.6^{\dagger}$ & $0.6^{\dagger}$ & $0.6^{\dagger}$ 
        \\
        \midrule
	\multirow{2}{*}{Diskbb}
        & $kT_{\mathrm{in}}(\mathrm{keV})$ & $0.640^{+0.011}_{-0.011}$ & $0.650^{+0.012}_{-0.006}$ & $0.784^{+0.003}_{-0.003}$ & $0.756^{+0.001}_{-0.001}$ & $0.739^{+0.001}_{-0.001}$ & $0.733^{+0.002}_{-0.002}$  \cr
        & $\mathrm{N_{Diskbb}}(\times10^{4})$ & $1.59^{+0.71}_{-0.16}$ & $2.00^{+0.10}_{-0.06}$ & $2.14^{+0.11}_{-0.03}$ & $2.22^{+0.02}_{-0.02}$ & $2.71^{+0.02}_{-0.02}$ & $2.83^{+0.02}_{-0.03}$ 
        \\
        \midrule
        \multirow{8}{*}{relxillCp}
        & $a_{*}$ & 
        $>0.78$ & 
        $>0.84$ & 
        $>0.77$ & 
        $0.86^{+0.04}_{-0.05}$ & 
        $0.84^{+0.07}_{-0.07}$ & $0.79^{+0.08}_{-0.08}$  \cr
        & $i(\mathrm{deg})$ & $25.1^{+3.1}_{-6.7}$ & $33.2^{+4.3}_{-4.5}$ & $28.0^{+1.2}_{*}$ & 
        $24.0^{+0.3}_{*}$ & 
        $23.0^{+0.3}_{*}$ & 
        $22.0^{+0.7}_{*}$  \cr
        & $\Gamma$ & 
        $2.04^{+0.02}_{-0.09}$ & $2.14^{+0.02}_{-0.04}$ & $2.25^{+0.04}_{-0.03}$ & $2.20^{+0.02}_{-0.01}$ & $2.21^{+0.02}_{-0.02}$ & $2.18^{+0.01}_{-0.02}$  \cr
        & $\mathrm{log}\xi$ & $3.24^{+0.57}_{-0.16}$ & $4.30^{+0.22}_{-0.38}$ & $3.70^{+0.74}_{-0.15}$ & $4.00^{+0.03}_{-0.02}$ & $4.00^{+0.06}_{-0.06}$ & $3.70^{+0.08}_{-0.36}$  \cr
        & $A_{\mathrm{Fe}}$ & $2.0^{+2.6}_{-0.4}$ & 
        $>4.8$ & 
        $3.8^{+4.1}_{-0.3}$ & 
        $>8.9$ & 
        $>9.0$ & 
        $>9.5$  \cr
        & $\mathrm{log}N$  & 
         $>19.65$ & 
         $<16.49$ & 
         $18.15^{+0.45}_{-1.51}$ & 
         $18.00^{+0.07}_{-0.08}$ & 
         $17.64^{+0.38}_{-0.26}$ & 
         $18.42^{+0.50}_{-0.21}$  \cr
        & $R_{\mathrm{ref}}$ & $0.73^{+0.20}_{-0.18}$ & $0.48^{+0.13}_{-0.10}$ & $0.81^{+0.26}_{-0.24}$ & $0.54^{+0.09}_{-0.07}$ & $0.67^{+0.17}_{-0.07}$ & $0.62^{+0.10}_{-0.05}$  \cr
        & $\mathrm{norm}$ & $0.13^{+0.01}_{-0.05}$ & $0.19^{+0.04}_{-0.02}$ & $0.09^{+0.03}_{-0.02}$ & $0.09^{+0.002}_{-0.003}$ & $0.06^{+0.004}_{-0.005}$ & $0.04^{+0.003}_{-0.004}$   \\
        \midrule
        \multirow{2}{*}{constant}
        & $C_{1}$ & 1.01 & 1.02 & 0.99 & 1.02 & 0.98 & 1.00  \cr
        & $C_{2}$ & 0.97 & 1.01 & 1.04 & 1.00 & 1.02 & 0.95   
        \\
         \midrule
        disk fraction & & 18.2\% & 23.2\% & 64.6\% & 64.6\% 
 & 77.9\% & 82.6\%
        \\
        \midrule
        $\chi^{2}/\nu$ & & 899.40/1285 & 941.97/1267 & 934.19/1248 & 1565.63/1285 & 1651.72/1285 & 1544.39/1285 
        \\
        \midrule
        $\chi^{2}_{\nu}$   & & 0.70 & 0.74 & 0.75 & 1.22 & 1.29 & 1.20
        \\
        
	\bottomrule
	\end{tabular}\vspace{0cm}
        \begin{tablenotes} 
		\item Notes: The best-fitting parameters obtained by \textit{Insight-HXMT} observations with model $\mathtt{constant*tbabs*(diskbb+relxillCp)}$. The parameters with the symbol “$\dagger$” indicate they are fixed at the values given. Symbol “*” indicates that the upper or lower limit of the parameter pegs the maximum or minimum value. The disk fraction is calculated by $\mathtt{cflux}$ in 2-20 keV.
     \end{tablenotes} 
	\label{tab:Training_sizes}
\end{table*}

\subsubsection{Markov Chain Monte Carlo Analysis}

\begin{figure*}
    \centering
    \includegraphics[scale=0.7,angle=-90]{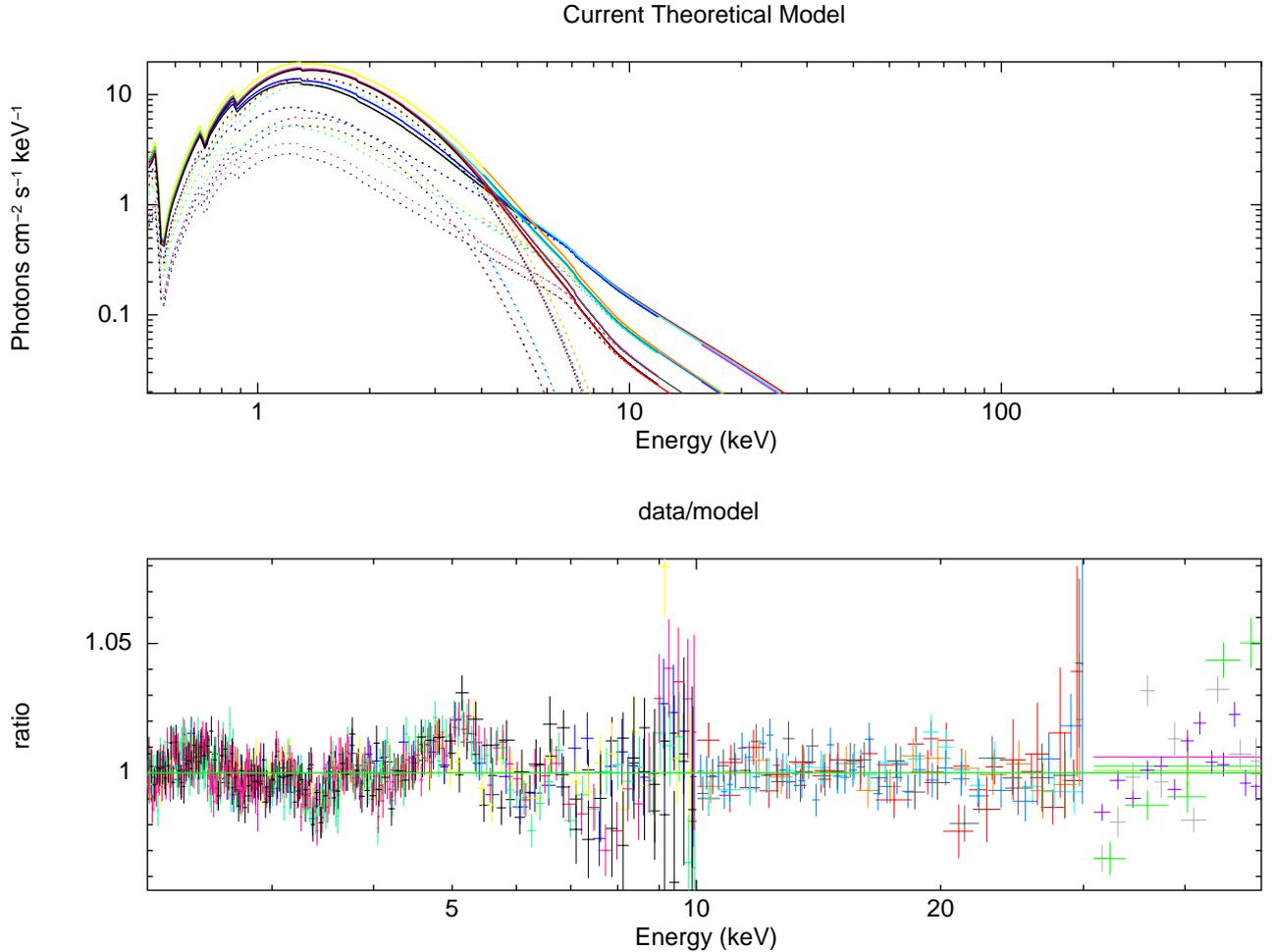}
    \caption{Top panel: best-fitting model (Model 6) for MAXI J1348-630. We fit 6 spectra simultaneously. Bottom panel: The ratio of fitting Insight-HXMT observations (model to data). The 6 spectra are well fitted, and the reduced chi-square $\chi^{2}/\nu$ of SP5-SP10 are concentrated around 1.}
    \label{fig:joint}
\end{figure*}

\begin{figure*}
    \centering
    \includegraphics[scale=0.7]{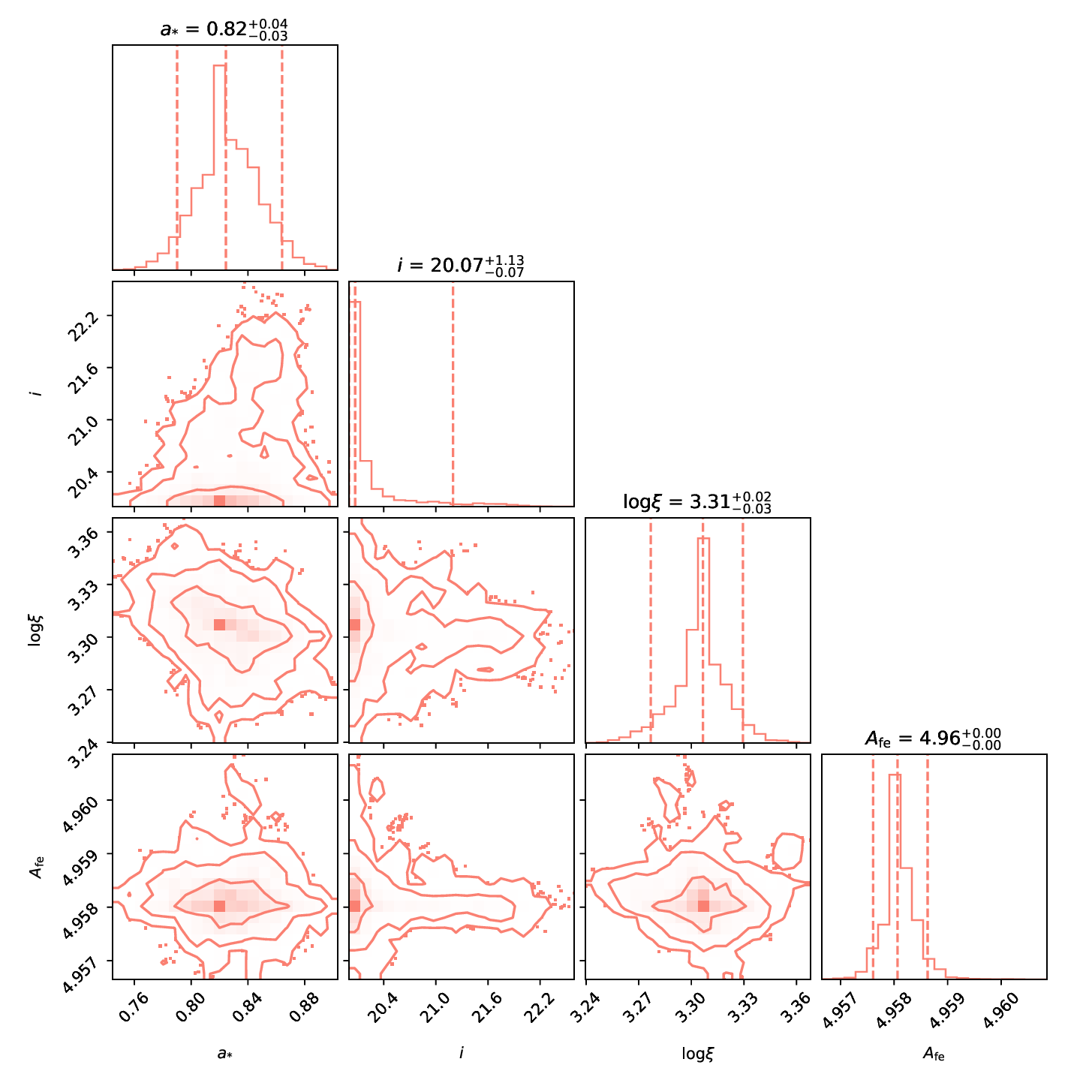}
    \caption{MCMC posterior probability distributions for joint fit with the corner package for the following parameters: black hole spin (dimensionless unit), inclination angle (in degrees), and ionization state of the disk and iron abundance (in units of solar iron abundance). The three contour lines represent 99.7\% (3$\sigma$), 95.4\% (2$\sigma$), and 68.3\% (1$\sigma$), respectively. The errors given above the panel are within the 90\% confidence level.}
    \label{fig:fit}
\end{figure*}

After fitting spectra with each of the models listed above, we jointed SP5-SP10 to analyze through the Markov Chain Monte Carlo (MCMC) in XSPEC (Model 6). For the fitting, we employ the Goodman-Weare algorithm with a number of steps of 100 and a total chain length of $10^{5}$ steps \citep{goodmanEnsembleSamplersAffine2010}. We discarded the first $10^4$ steps in the burn-in phase. When free the inclination to fit, it will always fall to the lower limit and be a poor fit. So we limited the inclination to between 20-40 degrees. Using the Geweke test, the convergence was determined by whether the absolute value of Geweke convergence is less than 2. And the effective sample size is 100000. In addition, we created time-series plots for all free parameters to check the convergence. The best-fit results for joint fit obtained by MCMC are listed in Table~\ref{tab:5}. In addition, since MCMC does not optimize the reduced chi-square, XSPEC obtains the best-fit value ($\chi^{2}/\nu$), while MCMC estimates the error. Figure~\ref{fig:joint} shows the best-fitting results given by the joint-fitting model. As shown in Figure~\ref{fig:fit}, we created MCMC contour plots for several parameters using the $\mathtt{corner}$ package \citep{corner}. Obviously, all these parameters can be constrained well. \\

\begin{table*}
	\centering
	\fontsize{8}{12}\selectfont    
	\caption{Joint-fit parameters with Model 6, inner radius were frozen at $R_{\mathrm{ISCO}}$ }
        \label{tab:5}
	\begin{tabular}{cccccccc}
		\toprule
		\multirow{3}{*}{\shortstack{Spectral\\ Component}}&\multirow{2}{*}{Parameter}&\multicolumn{2}{c}{HIMS}& \multicolumn{2}{c}{SIMS}& \multicolumn{2}{c}{HSS}  \cr
		\cmidrule(lr){3 -4} \cmidrule(lr){5-6} \cmidrule(lr){7-8}
		 && SP5 & SP6 & SP7 &SP8 &SP9 &SP10  \cr
        \midrule
        Tbabs & $N_{\mathrm{H}}(\times10^{22}\mathrm{cm}^{-2})$ & $0.6^{\dagger}$ & $0.6^{\dagger}$ & $0.6^{\dagger}$ & $0.6^{\dagger}$ & $0.6^{\dagger}$ & $0.6^{\dagger}$
        \\
        \midrule
	\multirow{2}{*}{Diskbb}
        & $kT_{\mathrm{in}}(\mathrm{keV})$ & 
        $0.627^{+0.008}_{-0.001}$ & 
        $0.664^{+0.008}_{-0.003}$ & 
        $0.779^{+0.004}_{-0.001}$ & 
        $0.755^{+0.003}_{-0.001}$ & 
        $0.739^{+0.002}_{-0.001}$ & 
        $0.732^{+0.003}_{-0.000}$   \cr
        & $\mathrm{N_{Diskbb}}(\times10^{4})$ & 
        $1.902^{+0.001}_{-0.125}$ & 
        $1.807^{+0.033}_{-0.103}$ & 
        $2.218^{+0.011}_{-0.052}$  & 
        $2.215^{+0.009}_{-0.043}$ & 
        $2.684^{+0.020}_{-0.018}$ & 
        $2.837^{+0.009}_{-0.052}$ 
        \\
        \midrule
        \multirow{8}{*}{relxillCp}
        & $a^{*}$ & \multicolumn{6}{c}{$0.82^{+0.04}_{-0.03}$}  \cr
        & $i(\mathrm{deg})$ &  \multicolumn{6}{c}{$20.1^{+1.1}_{-0.1}$}   \cr
        & $\Gamma$ & 
        $2.08^{+0.00}_{-0.01}$ & 
        $2.11^{+0.01}_{-0.00}$ & 
        $2.20^{+0.01}_{-0.00}$ & 
        $2.17^{+0.01}_{-0.00}$ & 
        $2.15^{+0.01}_{-0.00}$ & 
        $2.13^{+0.01}_{-0.00}$    \cr
        & $\mathrm{log}\xi$ & \multicolumn{6}{c}{$3.31^{+0.02}_{-0.03}$}   \cr
        & $\mathrm{log}N$  & \multicolumn{6}{c}{$>19.91$} \cr
        & $A_{\mathrm{Fe}}$ & \multicolumn{6}{c}{$4.9581^{+0.0006}_{-0.0005}$}  \cr
        & $R_{\mathrm{ref}}$ & 
        $0.26^{+0.02}_{-0.01}$ & 
        $0.24^{+0.04}_{-0.01}$ & 
        $0.47^{+0.08}_{-0.00}$ & 
        $0.51^{+0.07}_{-0.02}$ & 
        $0.80^{+0.10}_{-0.01}$ & 
        $0.80^{+0.12}_{-0.01}$  \cr
        & $\mathrm{norm}$ & 
        $0.187^{+0.003}_{-0.003}$ & 
        $0.186^{+0.003}_{-0.003}$ & 
        $0.089^{+0.005}_{-0.003}$ & 
        $0.077^{+0.003}_{-0.002}$ & 
        $0.040^{+0.002}_{-0.001}$ & 
        $0.033^{+0.002}_{-0.001}$ 
        \\
        \midrule
        \multirow{2}{*}{constant}
        & $C_{1}$ & 1.01 & 1.03 & 0.99 & 1.02 & 0.98 & 0.98   \cr
        & $C_{2}$ & 0.95 & 0.98 & 1.01 & 1.01 & 1.03 & 0.96    
        \\
        \midrule
        $\chi^{2}/\nu$ &  & \multicolumn{6}{c}{7555.07/7655}
        \\
        \midrule
        $\chi^{2}_{\nu}$   & & \multicolumn{6}{c}{0.99}
        \\
        
	\bottomrule
	\end{tabular}\vspace{0cm}
        \begin{tablenotes} 
		\item Notes: The best-fitting parameters obtained by joint-fit \textit{Insight-HXMT} observations with model $\mathtt{constant*tbabs*(diskbb+relxillCp)}$. The parameters with the symbol “$\dagger$” indicate they are fixed at values given. 
     \end{tablenotes} 
	\label{tab:Training_sizes}
    \end{table*}

\section{Discussion}\label{sec4}
\subsection{The Truncation Radius in Inner Disk}

This paper reports the analysis of data from the first outburst of MAXI J1638-430 observed by \textit{Insight-HXMT} in 2019. Spectra are fitted using a series of reflection models $\mathtt{relxill}$. The relevant basic parameters of this black hole transient source are obtained, see Table~\ref{tab:2}. The Fe K$\alpha$ fitting method is based on that the inner radius of the accretion disk extends to the ISCO, i.e. $R_{\mathrm{in}}=R_{\mathrm{ISCO}}$, and that the innermost accretion disk is not fully ionized. The canonical evolution of the accretion disk tells us that the disk will have been truncated in the LHS and would extend to the ISCO radius when it evolves into the HSS and the IMS. In order to test this phenomenon, we also made another fitting by fixing the spin parameter at the maximum value 0.998, and letting the radius of inner disk fit freely in the reflection model $\mathtt{relxillCp}$, see Table~\ref{tab:3}. However, the results show that the radius of the disk ($R_{\mathrm{in}}$) increases with time, which seems to be an anomalous behavior (Figure~\ref{fig:table3}). This result may be unphysical. A number of papers have used the blackbody-like flux as the true flux of the accretion disk, and from which the inner radius of the disk was inferred \citep{millerProminentAccretionDisk2006,zhaoEstimatingBlackHole2021}.  In the model $\mathtt{diskbb}$, $N_{\mathrm{disk}}$ is the normalization: $N_{\mathrm{disk}}=(R_{\mathrm{in}}/D_{10})^{2}\times \mathrm{cos\theta}$ \citep{kubotaEvidenceBlackHole1998}, where $R_{\mathrm{in}}$ is the apparent inner disk radius in units of km. Therefore, we calculated disk contribution using $\mathtt{cflux}$ function from XSPEC component to check the flux in 2-20 keV of $\mathtt{diskbb}$ component. In Table~\ref{tab:4}, we list the fraction of the disk component for all these spectra. It can be seen that the disk component was dominant during SIMS and HSS. As defined in \citet{kubotaEvidenceBlackHole1998}, the realistic inner disk radius is $r_{\mathrm{in}}=\xi\cdot f^{2}\cdot R_{\mathrm{in}}$, where $\xi$ is a correction factor for the stress-free inner boundary condition and $f\sim 1.7-2.0$ is the color correction factor. It seems that the higher $f$ in the HIMS explains the anomalously low values of $R_{\mathrm{in}}$ \citep{dunnGlobalStudyBehaviour2011}. Overall, the average inner radius of the six spectra is $R_{\mathrm{in}}=2.31\pm0.66$ leads us to believe that $R_{\mathrm{in}}$ have reached the ISCO in all six spectra.

\subsection{High Iron Abundance}

For SP5-SP7, the Model 4 fit gives nearly 5 times the iron abundance higher than the Solar abundance, while the SP8-SP10 gives about 10 times the abundance of the Sun (Table~\ref{tab:3}). A similar result was obtained by \citet{jiaDetailedAnalysisReflection2022} using \textit{NuSTAR} data analysis and \citet{kumarEstimationSpinMass2022} using \textit{NuSTAR} data and \textit{NICER} data. It should be noted that the measurement may not be representative of the true abundance of the system. The iron abundance of the super-solar system is a well-known problem in fitting relativistic reflection features: the large changes in the low energy band of the spectrum had an indirect effect on the iron abundance values measured by the spectral fit, and the value of iron abundance decreased when the electron density increased \citep{dongDetailedStudyReflection2020a,fengEstimatingSpinBlack2022}. So we free the $\mathrm{log}N$ for Model 5 to fit (Table~\ref{tab:4}). This model allows disk densitiy of up to $10^{20}\mathrm{cm^{-3}}$. Since the constraint on the inclination was not very good, we restricted the inclination to the same range as in the previous results. We found that after increasing the density of the disk, the iron abundance only showed a slight decline (Table~\ref{tab:4}). In addition, previous study results have also shown that its iron abundance is about $5Z_\odot$ \citep{mallBroadbandSpectralProperties2023c}. About the high disk density of MAXI J1348-630, the detailed discussions were held by \citet{chakrabortyNuSTARMonitoringMAXI2021}, they used the high-density reflection model $\mathtt{relflionx\_hd}$ \citep{tomsickAlternativeExplanationsExtreme2018b}, which has the disk density extending up to $\mathrm{10^{22}cm^{−3}}$. Using this model, they found the disk densities are more or less constant at $\sim \mathrm{10^{20.3-21.4}cm^{-3}}$ and the iron abundance is found to be close to the solar value. Furthermore, according to the histogram distribution of $A_{\mathrm{Fe}}$ obtained by \citet{garciaProblemHighIron2018}  from modeling the reflection spectra of 13 AGNs and 9 BHBs, the iron abundance peak around 5 appears to be a typical value for most sources. The large iron abundance required to fit the spectra by the reflection model is currently a problem and is still under investigation. Finally, we fixed the iron abundance at 5 and fitted it to obtain the spin parameter (Model 7). The results are shown in the table~\ref{tab:fe}. It can be seen that changing the $\mathrm{A_{Fe}}$ has little effect on the spinning results. In conclusion, the results we obtained for the spin value are consistent with the previous ones \citep{jiaDetailedAnalysisReflection2022,kumarEstimationSpinMass2022}, and all models with different components(Model 3, 5, 7) are in good agreement for spin value. So our final adopted spin parameter is not affected.

\begin{table*}
	\centering
	\fontsize{8}{12}\selectfont    
	\caption{Best-fitting parameters with Model 7, iron abundance is fixed at $A_{\mathrm{Fe}=5}$}
        \label{tab:fe}
	\begin{tabular}{cccccccccc}
		\toprule
		\multirow{3}{*}{\shortstack{Spectral\\ Component}}&\multirow{2}{*}{Parameter}&\multicolumn{2}{c}{HIMS} &\multicolumn{2}{c}{SIMS} &\multicolumn{2}{c}{HSS}  \cr
		\cmidrule(lr){3 -4}  \cmidrule(lr){5 -6} \cmidrule(lr){7-8} 
		 && SP5 & SP6 & SP7 & SP8 & SP9 & SP10  \cr
        \midrule
        Tbabs & $N_{\mathrm{H}}(\times10^{22}\mathrm{cm}^{-2})$ & $0.6^{\dagger}$ & $0.6^{\dagger}$ & $0.6^{\dagger}$ & $0.6^{\dagger}$ & $0.6^{\dagger}$ & $0.6^{\dagger}$ 
        \\
        \midrule
	\multirow{2}{*}{Diskbb}
        & $kT_{\mathrm{in}}(\mathrm{keV})$ & $0.609^{+0.015}_{-0.010}$ & $0.651^{+0.008}_{-0.007}$ & $0.781^{+0.001}_{-0.002}$ & $0.759^{+0.001}_{-0.002}$ & $0.741^{+0.001}_{-0.001}$ & $0.735^{+0.002}_{-0.002}$  \cr
        & $\mathrm{N_{Diskbb}}(\times10^{4})$ & $2.27^{+0.23}_{-0.39}$ & $2.00^{+0.10}_{-0.06}$ & $2.18^{+0.05}_{-0.04}$ & $2.22^{+0.02}_{-0.01}$ & $2.67^{+0.02}_{-0.02}$ & $2.80^{+0.03}_{-0.03}$ 
        \\
        \midrule
        \multirow{8}{*}{relxillCp}
        & $a_{*}$ & 
        $>0.79$ & 
        $>0.84$ & 
        $>0.76$ & 
        $0.85^{+0.04}_{-0.04}$ & 
        $0.73^{+0.07}_{-0.06}$ & $0.75^{+0.08}_{-0.09}$  \cr
        & $i(\mathrm{deg})$ & $20.0^{+5.5}_{*}$ & $35.0^{+7.2}_{-7.0}$ & $28.0^{+1.5}_{*}$ & 
        $24.0^{+0.2}_{*}$ & 
        $23.0^{+0.4}_{*}$ & 
        $22.0^{+0.7}_{*}$  \cr
        & $\Gamma$ & 
        $1.99^{+0.07}_{-0.04}$ & $2.16^{+0.01}_{-0.04}$ & $2.27^{+0.03}_{-0.03}$ & $2.18^{+0.01}_{-0.01}$ & $2.24^{+0.01}_{-0.01}$ & $2.17^{+0.02}_{-0.02}$  \cr
        & $\mathrm{log}\xi$ & $3.24^{+0.57}_{-0.16}$ & $4.14^{+0.09}_{-0.12}$ & $3.70^{+0.74}_{-0.15}$ & $3.30^{+0.02}_{-0.02}$ & $3.70^{+0.02}_{-0.02}$ & $3.06^{+0.11}_{-0.07}$  \cr
        & $A_{\mathrm{Fe}}$  & $5^{\dagger}$ & $5^{\dagger}$ & $5^{\dagger}$ & $5^{\dagger}$ & $5^{\dagger}$ & $5^{\dagger}$ \cr
        & $\mathrm{log}N$  & 
         $19.54^{+0.11}_{-0.88}$ & 
         $<17.04$ & 
         $17.46^{+0.66}_{-0.43}$ & 
         $19.74^{+0.17}_{-0.16}$ & 
         $18.00^{+0.11}_{-0.05}$ & 
         $>19.76$  \cr
        & $R_{\mathrm{ref}}$ & $0.80^{+0.48}_{-0.35}$ & $0.49^{+0.14}_{-0.10}$ & $0.69^{+0.33}_{-0.23}$ & $0.55^{+0.05}_{-0.04}$ & $0.69^{+0.06}_{-0.04}$ & $0.67^{+0.12}_{-0.13}$  \cr
        & $\mathrm{norm}$ & $0.11^{+0.05}_{-0.03}$ & $0.19^{+0.02}_{-0.04}$ & $0.11^{+0.01}_{-0.02}$ & $0.08^{+0.001}_{-0.002}$ & $0.06^{+0.002}_{-0.004}$ & $0.04^{+0.004}_{-0.005}$   \\
        \midrule
        \multirow{2}{*}{constant}
        & $C_{1}$ & 1.01 & 1.02 & 1.00 & 1.03 & 0.96 & 0.99  \cr
        & $C_{2}$ & 0.97 & 1.02 & 1.04 & 1.01 & 1.02 & 0.95   
        \\
        \midrule
        $\chi^{2}/\nu$ & & 903.72/1285 & 943.04/1267 & 934.38/1248 & 1582.77/1285 & 1676.92/1285 & 1576.51/1285 
        \\
        \midrule
        $\chi^{2}_{\nu}$   & & 0.70 & 0.74 & 0.75 & 1.23 & 1.30 & 1.23
        \\
        
	\bottomrule
	\end{tabular}\vspace{0cm}
        \begin{tablenotes} 
		\item Notes: The best-fitting parameters obtained by \textit{Insight-HXMT} observations with model $\mathtt{constant*tbabs*(diskbb+relxillCp)}$. The parameters with the symbol “$\dagger$” indicate they are fixed at the values given. Symbol “*” indicates that the upper or lower limit of the parameter pegs the maximum or minimum value.
     \end{tablenotes} 
	\label{tab:Training_sizes}
\end{table*}

\subsection{The Inclination Differencial } \label{4.3}
We note that the inclination of this black hole varies for different spectra. For SP5-SP7, they correspond to a higher inclination value, while for SP8-SP10 they are lower. As is known, accurately measuring the blue wing of the $\mathrm{Fe\ K}\alpha$ profiles is essential for constraining the inner disk inclination \citep{millerRelativisticXrayLines2007}. Though the line profile is clear, we don't get a good limit on the inclination for the $\mathtt{relxillCp}$ model when fitting the spectra separately. Apart from this, we note a slight decrease in $\Gamma$ and $\mathrm{log}\xi$ when shifting the model. This is due to the higher density producing additional soft X-ray emission, allowing for a harder direct power law \citep{tomsickAlternativeExplanationsExtreme2018a}. We found moderate
differences between our inclination results and the work of \citet{jiaDetailedAnalysisReflection2022} and \citet{kumarEstimationSpinMass2022}, but a bigger difference with \citet{titarchukMAXIJ13486302023}.  In addition, the existing estimates for this source have a large scatter in distance and black hole mass. Measurements using different methods varied widely. However, results using the same methods were close. \citet{tominagaDiscoveryBlackHole2020} found that the mass of the black hole is smaller when the spin and inclination are at low values. All of these estimates are based on a great number of model parameters. Perhaps a more refined approach to the determination of these system parameters will be necessary in the future.
 In brief, because we are more concerned with the spin, we do not over-interpretation of the inclination values.

\subsection{Variation of spectral parameters }
We have compared the parameter changes of SP5-SP10 in the case of Model 4 and Model 5. The changes in the ionization parameters and inner disk temperature follow the same trend as the changes in the luminosity for Model 4 (Figure~\ref{fig:table3}). In the HIMS, the inner disk temperature ($kT_{\mathrm{in}}$) from 0.617 keV increases to 0.780 keV. Then, it gradually decreases in the SIMS. The variation of temperature with luminosity is consistent with \citet{zhangPeculiarDiskBehaviors2022}. The ionization state of the disk is defined as $\xi=4\pi F/n_{e}$, where $n_{e}$ is the electron density of the disk \citep{fabianBroadIronLines2000a}. Our results from fitting the iron line profile follow the trend above. However, the ionization parameters $\mathrm{log}\xi$ do not follow the same trend in Model 5 because the disk density $\mathrm{log}N$ is also a free parameter (Figure~\ref{fig:table4}). So the variation of the ionization parameter is reasonable considering the disk density given by the fitting results. We can also see a decrease in the value of $\Gamma$ between Model 4 and Model 5. As mentioned in Section~\ref{4.3}, the decreasing trend is also related to the disk density, so for SP6, which has the least variation in the density of the disk, the value of $\Gamma$ also has the least variation.

\subsection{Low Reflection Fraction}
The reflection fraction ($R_{\mathrm{ref}}$) is defined as the ratio of the intensity of the primary source irradiating the disk and the intensity directly going to infinity \citep{dauserRelativisticReflectionRelxill}. The five models with different forms all showed low reflection fractions $R_{\mathrm{ref}}<1$. It is possible that this is due to the presence of a thick corona around the black hole that intercepts some of the thermal photons and Compton-scatters them. Additionally, the source can have a velocity away from the black hole, which also reduces the reflection fraction. \citep{steinerSelfconsistentBlackHole2017,steinerSTRONGERREFLECTIONBLACK2016}. 

\section{Conclusion} \label{sec5}

In this work, we have analyzed the archived \textit{Insight-HXMT} data of the black hole MAXI J1348-630 and investigated their corresponding spectral properties in the energy range 2.1-50 keV. We found six spectra show relatively strong reflective features, and therefore, using our adopted family of models, including the more physical relativistic reflectance model $\mathtt{relxillCp}$, we fitted all six spectra and obtained a good fit to all of them. This suggested that our disk radius in the intermediate state also stays around ISCO. We made a joint fit to all six spectra. In this case, the spin parameter is $a_{*}=0.82\pm^{0.04}_{0.03}$ at 90\% statistical confidence (statistical only). This indicates that MAXI J1348-630 has a moderate spin.

\section*{Acknowledgements}

 The authors thank the referee for the helpful comments to improve our manuscript. This work is supported by the NSFC (12273058). This work made use of the data from the \textit{Insight-HXMT} mission, a project funded by China National Space Administration (CNSA) and the Chinese Academy of Sciences (CAS).

\section*{Data Availability}

This article was completed using \textit{Insight-HXMT} data, which can be obtained from \url{http://hxmten.ihep.ac.cn/}.



\bibliographystyle{mnras}



\bibliography{ref}


\bsp	
\label{lastpage}
\end{document}